\documentclass[aps,showpacs,prl]{revtex4}
\usepackage{graphics}

\begin{document}

\title{\boldmath{Some (further) Comments on the $\Theta(1540)$ Pentaquark }}

\author{Shmuel Nussinov}
\email{nussinov@post.tau.ac.il}

\affiliation{School of Physics and
Astronomy, Tel Aviv University\\
Ramat-Aviv, Tel Aviv 69978, Israel\\ and \\
Department of Physics, University of Maryland\\
College Park, Maryland (USA) 20742-4111}

\begin{abstract}
Additional broader I=0 states in the KN channel near
$\Theta^+$(1540) are expected in many models, making the absence
of any signature in the K$^+$-deuteron scattering data even more
puzzling.

In an ideal ``three-body" picture the $\Theta$ is viewed as two
compact ud(1)ud(2)  $\bar{3}$ color diquarks and an $\bar{s}$
quark.    A ``QCD-type" inequality involving $m(\Theta^+), \;\;
m(\Lambda)$, the mass of  the $\Lambda(1/2^-)\;$ L=1 excitation
and that of a new I=0 tetraquark vector meson then follows. The
inequality suggests a very light new vector meson, and is
violated.

We note that ``associated production" of the pentaquark with
another quadriquark or anti-pentaquark may be favored. This along
with some estimates of the actual production cross sections
suggest that
 the $\Theta$ can be found in BaBar or Belle
e$^+$-e$^-$ colliders.
\end{abstract}

\maketitle

\subsection{I. Introduction and first comment}

The low mass and narrow width of the recently
discovered\cite{1Nakano}-\cite{8Chekanov} $\Theta$(1540)
pentaquark were anticipated to a certain extent in some chiral
soliton models.\cite{9Praszlowicz},\cite{10Diakonov} Yet, despite
the extensive research using more conventional QCD quark
models,\cite{11JW1}-\cite{17qm} these remain puzzling features.

This is particularly the case for the narrow width. The evidence
comes from two sources:  (i) directly from the measured width of
the $\Theta$ peak in the K$^+$n or K$^0$p invariant mass
distribution, and (ii) indirectly from the lack of evidence for
resonances in K$^+$d scattering data. The bounds on the width
from (i) cannot be lower than the
 resolution which in most experiments to date is no
better than O(10MeV);
\begin{equation}
   \Gamma{\Theta}| {\rm(direct \; observations)} \leq 10 MeV
\label{DirObs(1)}
\end{equation}
The indirect
bounds\cite{13N1},\cite{18Arndt-width2}-\cite{21Cahn-width5} use
the fact that KN elastic and  charge exchange resonant cross
sections have a Breit-Wigner shape with fixed normalization.
Hence, in older, low energy K-deuteron data, a  K$^+$N resonance
due to the $\Theta$ should manifest  as an  enhancement of the
cross section $\delta(\sigma)$. As long as the intrinsic
$\Gamma(\Theta)$ is smaller than the broadening due to the Fermi
motion of the neutron inside the deuteron, which is clearly the
case here, $\delta(\sigma)$ is proportional to $\Gamma(\Theta)$.

The analyses vary in sophistication and  the scope of
experimental information used suggesting:
\begin{equation}
  \Gamma(\Theta) \vert {\rm(indirect \; bounds)} < 1-3  MeV
\label{IB(2)}   
\end{equation}
To explain the small widths we need quark (or other) models where
the $\Theta$ has a small overlap with the KN decay channel.

It was recently suggested\cite{22KL3} that mixing of  (1/2)$^+$
states via the KN continuum yields  physical states of which
one---the $\Theta$(1540)---is anomalously narrow while the  other
states have ``normal", say, 50-200 MeV widths. Additional states
in the $ I=0 $ KN  channel in the 300 MeV range above the
$\Theta$(1540) are generic. Indeed the spectrum in the more
complex five-quark system is expected to be denser than in the
corresponding non-exotic three quark baryonic system. It is
instructive to see how these general expectations are realized in
specific models. In the model\cite{14KL1},\cite{15KL2} with one
(ud) diquark and an  ($\bar{s}$ud) compound, the $\bar{s}$
tunnels between the two ud diquarks. This yields a lowest
symmetric (1/2)$^+$ state and a slightly higher antisymmetric
(1/2)$^-$ state. In the two-diquark model, the two I=S=0 diquarks
are in a relative orbital angular momentum,
L(12)=1\cite{11JW1}-\cite{13N1}. Coupling this orbital angular
momentum with the spin-half of the
 $\bar{s}$ antiquark yields the J=1/2 $\Theta$ and a
slightly higher J=3/2 state. The $\vec{L} \cdot \vec{S}$ (spin
orbit) splittings in ordinary hadrons suggest
 a  splitting of less than 250 Mev between these two
states.\cite{16JM}
 We assumed all along that the $\Theta$ is a positive
parity spin-half particle.
 However, lattice and some early quark model
calculations\cite{16JM} suggest that we have
 all quarks in an S wave yielding a $(1/2)^-$ state.
Thus we could also have a splitting smaller than 300 MeV between
the two states of opposite parity.

A trivial yet key observation is that the indirect bounds (ii)
apply not  only to the  putative $\Theta(J(P=1/2^+))$, but to {\it
any} resonance in the KN channels.   Hence, not only is
$\Gamma(\Theta)$ constrained, but also the width of any  other I=0
state which couples to the KN channel in the mass range of
1540-18000 MeV. (I spin forbids the decay of such states into
$\Theta$(1540) and a pion maintaining a small inelasticity).
These states could be the extra, broader (1/2)$^+$ states
invoked\cite{22KL3} to explain the narrow width of $\Theta$ or
any one of the higher states discussed above. Thus, let us
consider the $\Theta(3/2)^+$. The resonant cross section is:
$\sigma$(KN) (res) $\propto$ (2J+1)/p(cm)$^2$. The higher 3/2 spin
makes it roughly the same as the resonant KN  cross sections due
to $\Theta$ J=1/2 at a lower p(cm). Hence, the width of a J=3/2
state in the ~300 MeV mass range above the $\Theta$   should,
like that of the $\Theta$(1540), be bound by
\begin{equation}
  \Gamma(\Theta(J(P)=(3/2)^+) < 1-3 MeV
\label{NIB(3)}     
\end{equation}

This new indirect bound (NIB) is even more problematic than the
original IB on $\Gamma(\Theta$ J=1/2): First, the centrifugal
barrier factor, $\sim$ p(cm)$^3$, suggests that the higher
$\Theta(3/2^+)$ state is much broader. Second, the $\Theta
(1540)$ corresponds to P[K](Lab)~440 MeV  near the lower limit of
all past measurements, whereas the higher states would manifest at
higher kaon momenta where more detailed and accurate data from
several different K$^+$-d experiments are available.

{\it Very} broad, say, $\Gamma > 200$ MeV, resonances may escape
detection in experiments looking for bumps in invariant (missing)
K-N mass distributions. However, these I=0 resonances reflect in
K$^+$d scattering. The total cross section over the relevant of
(p(Lab)= 500-900 MeV) range should increase from $\sigma$(K$^+$d)
$\sim$ 2$\sigma$(K$^+$)p $\sim$ 25 mb by the huge amount of
$\sigma$ (resonance) $\sim$ 30 mb. Thus the impact of the K$^+$d
data on pentaquark modeling is far-reaching:
 Not only should the $\Theta(1540)$ be very narrow,
similar bounds
 apply to the width of {\it any} I=0 K-N resonance in
the 1540-1800 MeV region.

\subsection{\boldmath{II. Second comment: A ``strict" point-like
diquark-diquark $\bar{s}$
 picture of the $\Theta$ conflicts with ``QCD
inequalities"}}

 The simple appealing model of the $\Theta$(1540) as
two I=S=0 ud diquarks carrying a $\bar{3}$(c) of color each and
an $\bar{s}$ \cite{11JW1,12JW2,13N1} maximizes the attractive HF
interactions between the lighter, nonstrange quarks. The two
diquarks with $\bar{3}$(c) couple anti-symmetrically in color to
a 3(c) so as to form an overall color singlet with the remaining
$\bar{s}$ antiquark.  Bose statistics for the spinless identical
diquarks then implies an odd relative orbital angular momentum,
L(1,2), between the two diquarks so as to ensure overall
symmetry. The lowest odd angular momentum, L=1, coupled with the
spin-1/2 of $\bar{s}$ to the likely lower energy J=1/2 state,
yields a $\Theta$(1540) with spin/parity (1/2)$^+$ just as in the
chiral soliton models.  The centrifugal barrier due to the
orbital angular momentum also helps understanding the small
$\Theta$ width. (``tensor'' diquarks with 6(c) of color were
considered in Ref. \cite{23SZ}). However, the L=1 relative angular
momentum generates extra kinetic energy and the dq(1)
dq(2)$\bar{s}$ configuration may not be as light as 1540 MeV.  We
next argue that this is indeed the case if we treat the diquarks
as elementary scalars carrying 3$\bar{c} $ color.

The $\Theta$ pentaquark is then analogous to an (anti-) baryon
with two among the three antiquarks replaced by scalar diquarks,
dq(1) and dq(2), with the same 3$\bar{c}$ color. The pair-wise
interaction among the three colored objects in the baryon or in
the idealized pentaquark model has two parts with $\vec{\lambda}^i
\cdot \vec{\lambda}^j$ (or $1^i \cdot 1^j)$ color structure.
 We will assume that the first class of interactions
dominate (as is the case in  the large N$_c$ limit). As indicated
in Ref. \cite{24N2}, we can then generate  the three-particle wave
function---by charge conjugating one particle at a time---trial
wave functions for three mesonic two-body states. The variational
principle yields inequalities, between $m^{(0)}[B_{(i,j,k)}]$ and
$m^{(0)}[M_{(i,\bar{j})}]$, etc., the masses of the lightest
baryon mesons with the specific i,j,k flavors indicated are:
\begin{equation}
2\cdot m^{(0)}[B_{(i,j,k)}]  \geq
m^{(0)}[M_{(i,\bar{j})}]+m^{(0)}[M_{(j,\bar{i})}]+m^{(0)}[M_{(j,\bar{k})}]
\label{BMI(4)}       
\end{equation}
Rigorous inequalities $ m(N) > m(\pi)$ \cite{25Weingarten} and
$m(N) > 3/2 m(\pi)$ \cite{26Cohen} were obtained using the path
integrals for Euclidean qqq(x)qqq(y) (baryonic) and
$q\bar{q}(x)q\bar{q}(y)$ (mesonic) correlators. Table (ii) in the
recent review of QCD inequalities\cite{27NL} shows that the
above more detailed flavor-dependent inequalities hold with a
substantial 200-500 MeV margin in all testable cases.

The same reasoning yields in the present case:
\begin{equation}
2\cdot m(\Theta(1540)) = 2\cdot m^{(0)} (ud,ud,\bar{s}) \geq
a\cdot m(\Lambda_s(1/2)^+) + (2-a)\cdot m(\Lambda_s(1/2)^-) +
m^{(0)}[(ud)(\bar{u}\bar{d})(J(P)=1^-)]
\label{NPI(5)}  
\end{equation}
The baryon meson mass inequalities can also be derived in a QCD
string/flux
 tube picture for the baryons and mesons with just one
junction point in
 the baryons.\cite{24N2},\cite{27NL} A similar
description with two
 $(J\bar{J})$ and three $(J,J,\bar{J})$ junction
points and minimal 3(c)
 (or $\bar{3}$(c)) chromoelectric flux lines has been
suggested for a
 putative exotic tetraquark and the pentaquark.
 \cite{28CN},\cite{29GN}. It is straightforward to
verify that the
 derivation of the new inequality holds in such a
scenario as well.

The Lorentz quantum numbers of the states on the
right-hand side
of  Eq. (\ref{NPI(5)}) are inferred from the
pentaquark state.
The I=S=0
 ud(1) and ud(2) diquarks inside the $\Theta$ have
relative orbital angular momentum
 L(1,2)=1. Hence, also in the tetraquark state
appearing at the end of
 Eq. (\ref {NPI(5)}) we have angular momentum L=1
between the I=0,S=0
 diquark and anti-diquark making an I=0 vector meson.
The lightest I=0
 vector mesons are the well-known $\omega$(780)and
$\phi$(1020).  A
 tetraquark vector meson lighter than $\phi$(1020)
would have been detected
 in $e^+e^- \rightarrow 3 \pi$ reactions.  Hence, the
last $m^{(0)}[(ud)(\bar{u}\bar{d})(J(P)=1^-)] \equiv
 m[Tq(1^-)]$ value in Eq. (\ref{NPI(5)}) is {\it larger}
than 1020. Also,
 L[dq(1),dq(2)] = 1 excludes having both dq(1) and
dq(2) in an L=0 state
 relative to the remaining $\bar{s}$. We find that the
probability of having inside
 the $\Theta \;$ dq(1) - $\bar{s}$ (and independently,
dq(2) - $\bar{s}$) in an L=1 state exceeds 50\%, i.e.,
$a < 1$ in
Eq. (\ref{NPI(5)}) above.  Using
 $m(\Lambda)$ =1115 MeV and 1405 MeV for the mass of
$\Lambda(1/2)^-$---the
 first negative parity L=1 excitation---we find that
the inequality is strongly violated: $2 \cdot 1540$ MeV is {\it
smaller} than (1115+1405+1020) MeV by 450 MeV rather than larger
by $\sim$ 200-400 MeV, the margin with
 which the meson baryon inequalities were satisfied.

Obviously this does {\it not} exclude a light pentaquark
$\Theta$(1540).  Rather the extreme version of the two diquark
model may be wrong. Treating the diquarks as idealized point-like
scalars omits significant hyperfine attractive interactions
between the $\bar{s}$ and each of the four quarks.  (The latter
were incorporated via constructing the $u-\bar{s}-d$ aggregate in
Refs. \cite{14KL1},\cite{15KL2}).

Still, the above suggests that  a light pentaquark is accompanied
by light (crypto) exotic $dq-d\bar{q}$ tetraquarks.

Heavier flavor analogs of $\Theta$ (1540) with $\bar{c}/\bar{b}$
replacing $\bar{s}$ were considered by many authors. The smaller
HF interactions with the heavier $\bar{Q}$ suggest that the
idealized $dq(1)dq(2)\bar{Q}$ picture and the analog of the
inequality \ref{NPI(5)} are more applicable here. The latter
reads:
\begin{equation}
  2m(\Theta(c)) > m[\Lambda(c)1/2^+] +
m[\Lambda(c)1/2^-] + m[Tq (1^-)]
\label{NPcI(6)}           
\end{equation}
with $Tq(1^-)$ the vector tetraquark state encountered above.
Using 2.285 GeV for the mass of $\Lambda(c)$, 2.593 GeV for its
L=1 excitation, and $m Tq(1^-) >$1020 as above we find that
$m(\Theta_c) > 2.95 > m_D +m_N = 2790$. Thus $\Theta (c)$ is
likely to be unstable decaying
 into a D meson + Nucleon.

Possible further difficulties encountered in explaining N$^*$
widths in an SU(3) flavor  extended version of the $dq dq
\bar{q}$  picture have also been  noted.\cite{30Cohen} This is
hardly surprising as the dq(1)dq(2) picture is even more
questionable if we replace the $\bar{s}$ by even lighter
$\bar{u}/\bar{d}$.

\subsection{\boldmath{III. Third comment: ``Associated
Production of Exotics" and some crude
 estimates of $\Theta$ (1540) production in $e^+e^-$
colliders}}

 The $\Theta$(1540) pentaquark has been discovered in
many
 experiments. The early searches in K$^+$-deuteron
mentioned in Sec. I
 and $e^+e^-$ machines are two notable exceptions.

Recently  new charmed states  spectroscopy and a
 crypto-exotic $c\bar{c}q\bar{q}$ mesons have
been discovered in BaBar and Belle.  These $e^+e^-$ accelerators
primarily investigate CP violation and b-quark physics. However,
with  4$\pi$ coverage, precise momentum measurements and good
 particle identification, the O(10$^9$) events
involving lighter primary
 quarks collected there are a unique ``hadronic
treasure trove".

For $\Theta (\bar{s}uudd)$ production the $e^+e^-$ initial state
is the opposite of K$+$n collisions where {\it all} the requisite
four quarks and $\bar{s}$ are initially present. To produce
$\Theta$ (or $\bar\Theta$) in
\begin{equation}
   e^+ e^- \rightarrow \Theta + X
\label{STP(6)}       
\end{equation}
we need to create $udud$ and $\bar{s}$ and their anti-particles.
At the BaBar/Belle cm energy of W=$\sim$11 GeV many quark pairs
are eventually produced.   However, $\Theta$ is an aggregate of
{\it specific} five quarks within O(Fermi) spatial proximity and
with small relative momenta. Thus the production rate can be
strongly suppressed reducing the sensitivity of searches for
$\Theta$ in $e^+e^-$ collisions.

This seems to be even more the case for ``double", i.e.,
$\Theta\bar{\Theta}$ pair production:
\begin{equation}
   e^+e^- \rightarrow  \Theta + \bar{\Theta} + X
\label{DTP(7)}             
\end{equation}
Our third comment is that this need not be the case. The
$\Theta(1540)$ is the lightest member of a novel family of exotic
hadrons.  Once being produced, the latter hadrons will often
decay into the ground state $\Theta$. We next  argue that joint
``associated" production of the new exotics is likely and the
fraction of all hadronic events with
 double $\Theta$  production  (DTP) far exceeds the
square of the
 fraction of single $\Theta$ production events (STP).

Building the complicated pentaquark structure with all the
specific quarks and their correlations starting with the hadronic
vacuum and using the local flavor and color conserving QCD
interactions, we simultaneously produce a ``mirror" set of
antiparticles with the same color, spin and space, etc.,
correlations.  Both sets will, in general, yield very excited
pentaquark configurations, say, of the above $dq dq \bar{s}$
form.  If de-excitation occurs via pion or the $\eta$ (but {\it
not} via kaon emission!)---which can happen in a substantial
fraction of the cases---the ground state $\Theta$(1540) stays in
the final state. Thus while, as we show below, single $\Theta$
production (STP) is heavily suppressed, the even more striking
events where we have {\it two} $\Theta$'s (DTP) produced in the
same $e^+e^-$ collision may be only moderately further suppressed
relative to (STP).

We next present some rough estimates of these rates. At $W \sim
10$ GeV in $e^+ e^-$ collisions, the initial hard short-distance
process yields mainly just a quark and anti-quark ($u\bar{u}$ and
$d\bar{d}$ in about half the cases). We will use the
chromoelectric flux tube model developed for the soft
hadronization process in this case.\cite{31CNN} In this model a
constant chromoelectric field is generated inside a tube of fixed
diameter between the the separating initial quarks. Tunneling
within this field yields extra $q\bar{q}$ pairs and eventually
multiple mesons. The tunneling rate is proportional to $E^2
exp(-(\pi m^2)/g_{s}E)$ with m the quark mass  and E the constant
chromoelectric field.
 To  produce a $\Theta$ pentaquark specific
deviations from the generic scenario for multiple meson
production are required: Focusing on the quark side we need first
to produce a d or u quark
 which couples with the primary hard quark to a $\bar{3}$
of color. This is suppressed relative to the standard
$\bar{q}$  production yielding a color singlet meson. The
suppression traces to the fact that the effective E field is only
half as large as the one for meson production.
 The same argument implies that in the subsequent step
producing a $\bar{3}$(c) quark yielding
 a color singlet baryon is favored. The $\epsilon=
O(1/10)$ suppression of baryon production is
 thus a measure of the penalty incurred in the
unfavored di-quark production in the first step. However, building
the  $\Theta$  requires again  in the second step the production
of $\bar{s}$ so as to make a $ud\bar{s}$ cluster with overall
$\bar{3}(c)$ color. The production of such an $\bar{s}$ is
suppressed not only by $\epsilon$ due to the reduced field, but
due to the higher $s$ mass the overall suppression factor is
$\epsilon/9 \sim 1/100$.
 The chromoelectric field driving the tunneling here
has half the value as in  multi-meson production.
  The suppression due to $m^2_s > m^2_u$ is here 9;
the square of the factor 3 suppression
 for s quark production relative to u/d in the usual
multi-meson production case,

The last two steps of generating an excited $\Theta$-like state by
creating another ud and
 du quark are analogous to those in baryon production
 with an $\epsilon ~1/10$ suppression.
Collecting all factors we expect one ``Primary'' pentaquark
production event in $2\cdot(\epsilon)^{-3} \cdot 9 \propto 2 \cdot
10^4 \; e^+ e^-$ collisions. The primary excited pentaquark may
decay via kaon emission to some excited $N^*$, in which case we
will not find the $\Theta(1540)$ in the final state. However, we
expect that for an appreciable fraction, $f \propto 0.5$, of all
primary pentaquarks this will not be the case and:
\begin{center}
Probability of $ \Theta(1540)$ production $\sim 5 \cdot
10^{-5}\cdot f \sim 2.5 \cdot 10^{-5}$.
\end{center}
A key observation is that the probability of producing two
$\Theta(1540)$ particles in the same event is not the square of
the last small probability. Because our production mechanism
naturally yields two excited pentaquarks, all we need is for both
to decay into the ground state pentaquark and therefore:
\begin{center}
Probability of $\Theta(1540) + \bar{\Theta}(1540)$ production
 $\sim 5 \cdot 10^{-5} \cdot f^2 \sim 10^{-5}$.
\end{center}

 Since the $\Theta(1540)$ decays into $K^0p$  in half
the cases, and we
 further have a 1/3 reduction from the demand that the
$K^0$ will decay as
 $K_S$ into charged (rather than neutral pion pair)
the requirement of
 detectability reduces $f$ to an ``effective $f$" $\sim f/6$.
 This suggests a probability of $\sim 2\cdot 10^{-6}$ for
detecting a $\Theta$(1540) per collision.

 While these probabilities are pretty small, the large number
of collisions may
 still yield a significant signal.
We also note that if the $\Theta(1540)$ signal at Zeus and, in
particular, the
 appearance of approximately one hundred $\bar{\Theta}(1540)$ (which cannot be
proton fragments
 anyway) in four million e-proton collisions are
real,
 then the last probability is, in fact,
 $2.5 \cdot 10^{-5}$---consistent with our estimate!

The CFT model can be applied to hadron production and to
 pentaquark production, in particular, also off hadronic targets.

Thus in $\gamma$ proton collision we can consider the struck u,
say, quark separating from the ud
 diquark instead of the initial state of,
 say, $\bar{u}-u$ in $e^+ e^-$ collisions.

Following similar steps as above we see that the production of the
pentaquark requires one less step and, hence, we gain a
$1/\epsilon \sim$ factor 10 in production strength so that we
expect $\Theta(1540)$ to be produced in 10$^{-4}$ of all cases.
Also, the ``mirror" exotic that will be generated in this case
with the pentaquark will be a strange tetraquark
$(\bar{s}\bar{u})(ud)$.

We note that neither expectation seems to be born out: first the
accompanying hadrons were the ordinary non-exotic kaon, or K$^*$.
Finally, the pentaquark production rate is considerably larger in
this case than our estimates.

\subsection{Acknowledgements}

I would like to thank Marek Karliner for asking me some time ago
if QCD inequalities can be applied to the pentaquark, and to Ralf
Gothe for helpful discussions.

I enjoyed discussing aspects of pentaquark physics with Tom Cohen.
I did not, however, take his (sound!) advice and write a special
purpose paper with only one single motif---running the risk that
many readers will note, as in the first case\cite{13N1}, only one
of the three comments...

\end{document}